\documentstyle[twoside,fleqn,espcrc2,epsfig]{article}

\newcommand{\AmS}{{\protect\the\textfont2
           A\kern-.1667em\lower.5ex\hbox{M}\kern-.125emS}}

\title{ALTAI: Computational code for the simulations of TeV air showers 
       as observed with the ground-based imaging atmospheric \v{C}erenkov  
       telescopes.}

\author{A.K. Konopelko\address{Max-Planck-Institut f\"{u}r Kernphysik, 
                               Heidelberg D-69029, Postfach 10 39 80, Germany}%
  \thanks{Corresponding author. E-mail: Alexander.Konopelko@ 
                                   mpi-hd.mpg.de}
        A.V. Plyasheshnikov\address{Altai State University, Dimitrov 66, 
                                    Barnaul 656099, Russia}}

\begin{document}

\begin{abstract}
Ground-based atmospheric \v{C}erenkov telescopes are proven to be
effective instruments for observations of very high energy (VHE) 
$\gamma$-radiation from celestial objects. For effective use of 
such technique one needs detailed Monte Carlo simulations of 
$\gamma$-ray- and proton/nuclei-induced air showers in Earth 
atmosphere. Here we discuss in detail the algorithms used in the 
numerical code ALTAI developed particularly for the simulations of 
\v{C}erenkov light emission from air showers of energy below 50~TeV. 
The specific scheme of sampling the charged particle transport in
the atmosphere allows the performance of very fast and accurate simulations used 
for interpretation of the VHE $\gamma$-ray observations.      
\end{abstract}

\maketitle

\section{Introduction}
Recent exciting detections and observations of TeV $\gamma$-ray emission from a number of 
galactic and extragalactic objects (Ong 1998, Catanese \& Weekes 1998) have shown the high 
performance of currently operating imaging air \v{C}erenkov telescopes (IACTs). Several 
projects for future detectors have been proposed lately. The major forthcoming instruments, 
such as CANGAROO~IV, H.E.S.S., MAGIC and VERITAS, will have significantly better 
sensitivity to $\gamma$-ray fluxes in the energy range from 50~GeV up to 50~TeV.  

For a lack of a collimated test beam of TeV $\gamma$-ray photons, the ground-based 
\v{C}erenkov detectors heavily rely on the Monte Carlo simulations of the \v{C}erenkov 
light emission from air showers which are used to understand the performance of detector. 
Basic parameters of the instrument (detection area, angular and energy resolution, efficiency of 
cosmic ray rejection, etc) can be derived from the simulations. The crucial point is a 
precise measurement of the primary shower energy. 
For that one should include properly into the simulations all processes of \v{C}erenkov 
light emission in the air shower as well as photon propagation on the way from the 
emitting shower particle to the telescope camera. Measurements of $\gamma$-ray fluxes, 
energy spectra, upper limits strongly depend on the absolute energy calibration of a telescope.  

Here we give a description of ALTAI 
\footnote{ALTAI is the abbreviation for {\it Atmospheric Light Telescope Array Image}. 
Mountain {\it Altai} is the pristine wilderness in the south-west of Russia.}
computational code developed for detailed Monte Carlo 
simulations of the \v{C}erenkov light emission in TeV air showers. Among the other existing 
codes intended for such simulations, MOCCA (Hillas 1979), KASCADE (Kertzman \& Sembroski 1994), 
CORSIKA (Heck et al. 1997), this code has a distinctive advantage -- it's rather high speed of 
shower simulations due to a specific algorithm used for processing the multiple scattering of 
charged shower particles. This approach does not consume a lot of CPU time and 
allows to perform fast and accurate simulations. Together with the additional routine 
recently developed for the simulations of telescope response (Hemberger 1998) the ALTAI code was 
extensively used for production of a standard Monte Carlo database used in the HEGRA (High 
Energy Gamma Ray Astronomy) VHE $\gamma$-ray experiment (Konopelko et al. 1999a). 

The ALTAI code consists of two major programs which simulate the development of 
the electromagnetic (EMCCS) (hereafter we put in brackets the name of the corresponding 
subroutine of the code) and proton-nuclei (STEPAD, MULTIP, XPI) cascade in Earth 
atmosphere. We discuss in detail the procedure of simulating the electromagnetic and 
hadron-nuclei cascade in Sections~2 and 3, respectively. Section~2 also describes the 
algorithms of \v{C}erenkov light emission by the shower charged
particle. Section 4 deals with nucleus-nucleus interactions.  
In Section~5 we review the results of the \v{C}erenkov light simulations using ALTAI 
code compared with other relevant simulations and observational data.
\begin{figure}[t]
\includegraphics[width=1.00\linewidth]{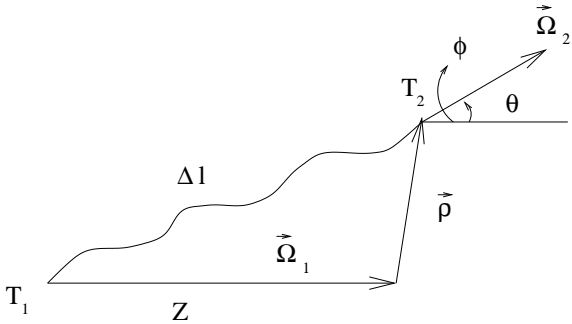}
\caption{Transformation of phase coordinates of a charged particle at the multiple 
scattering segment.}  
\label{fig:1}
\end{figure}

\section{Electromagnetic cascade}

Here we overview the part of the code intended for the generation of an electromagnetic 
cascade in the atmosphere (EMCCS). Note that at present all cross-sections for electron 
and photon interactions are established and are well described in detail in the relevant 
energy range. An electromagnetic part of the code accounts for the following interaction 
processes: electron-positron pair production, Compton  scattering for the primary photons 
and bremsstrahlung, ionization losses, Coulomb scattering for the primary electrons and 
positrons.

The cross-section of $e^-,e^+$ pair production by primary photon was calculated according to 
the Bethe-Heitler formula taken from Motz et al. (1964) (MPRH). The emission angle of $e^-,e^+$ 
components, defined with respect to the direction of initial photon, was sampled using the Bethe 
distribution (Motz et al. 1964). We used the Klein-Nishina formula (see e.g., Gaisser 1990) 
for the cross-section of Compton photon scattering (COMPH). The dependence of photon 
cross-sections on energy and the atomic number of a medium was derived from data taken 
from Storm \& Israel (1973) and Hubbel (1969) (FMPM).

We simulated the bremsstrahlung interaction process (MRDH) according to the Bethe-Heitler 
formula (see Koch \& Motz, 1959). The angle of emitted photon was sampled from the Schieff 
distribution (see Koch \& Motz, 1959). The Bethe-Bloch formula, with
the corrections for the density effect (Sternheimer 1952), 
was used for simulation of the mean ionization losses (MIONH) whereas 
for the differential cross-section of $e^-,e^+$ ionization collisions we used the M\"oller 
formula (M\"oller 1932). Note that in the electromagnetic shower positrons were treated 
similarly to electrons. 

\subsection{Multiple scattering}

The relativistic electron suffers an enormous number of interactions along its path length 
in the matter. Apparently, a direct simulation of all interactions is not possible. That 
is why all numerical codes for the shower simulation use a specific technique for grouping 
electron interactions (see for review Berger 1963, Akkerman 1991, Gaisser 1990). In this 
approach one can simulate only the so-called {\it catastrophic} $e^-,e^+$ interactions. 
These interactions lead to emission of photon/electron of a relatively high energy, 
$\rm E>Q$, which exceeds the intermediate energy threshold of $\rm Q$. The same 
approach was used for the simulation of the $e^-,e^+$ ionization process 
($\rm Q=Q_{I}$) 
and bremsstrahlung ($\rm Q=Q_{R}$). 
The emitted electron must have a kinetic energy above a certain intermediate energy threshold, 
$\rm T > Q_{I}$.  

A random path length between neighboring catastrophic collisions was divided into small 
segments of $\rm \Delta l \le 5\, gr/cm^2$. We sampled the phase coordinates of a charged 
particle at the end of the segment $\rm \Delta l$ according to the cumulative effect of 
all low energy collisions along the segment (SGMEH, SGMQH). Thus we simulated at the end 
of each segment the loss of the $e^-,e^+$ kinetic energy, $\rm \Delta T = T_1 - T_2$, the 
scattering angle, $\rm \Theta=\arccos( \vec{\Omega}_1\vec{\Omega}_2)$, the azimuth angle of 
scattered particle, $\rm \varphi$, the longitudinal $\rm Z$ and lateral $\rm \vec \rho$ 
displacement of initial particle while passing over the segment (see Figure~1). 
Appropriate transport equations were used in order to derive the probability 
distributions for the phase coordinates at the end of a segment. We have obtained the analytical 
solutions for these distributions.  

\subsubsection{Energy losses}

The energy losses of a charged particle, $\rm \Delta E$, in the segment, $\rm \Delta l$,
were sampled according to the Landau-Vavilov formula (Landau 1944, Vavilov 1957).
Plyasheshnikov \& Kolchuzkin (1975) have tabulated this formula for a specific 
conditions of shower simulations. The distribution of the energy losses at the 
end of a multiple-scattering segment was described as follows:
\begin{eqnarray}
P(\Delta T)d\Delta T = f(\mu,\lambda)d\lambda, \nonumber  \\
f(\mu,\lambda) = \frac{1}{2\pi i}\int_{c-i\infty}^{c+i\infty} \exp \left\{ \lambda t
- \mu \phi(t)\right\} dt, \nonumber \\
\phi(t) = t \int^t_0 \frac{(1- \tau - exp(-\tau))}{\tau^2} d\tau \nonumber \\
\mu=\Sigma_{h}(T_1)\Delta l, \nonumber \\
\lambda=(\Delta T - \beta_l (T_1)\Delta l)/Q_{I},\nonumber  \\
\Sigma_{h}(T)=\int_{Q_I}^{T/2}W_I (T,Q)dQ,
\label{eq11}
\end{eqnarray}
where  
\begin{eqnarray}
\beta_l (T)= \beta_I + \beta_R \nonumber \\
\beta_I = \int_0^{Q_I}W_I(T,Q)QdQ \nonumber \\
\beta_R = \int_0^{Q_R}W_R(T,Q)QdQ
\label{beta}
\end{eqnarray}
is the mean energy loss per unit path length due to the inelastic collisions of a charged 
particle by production of low energy secondaries. $\rm W_I(Q), W_R(Q)$ are the differential 
cross-sections for the ionization ($\rm I$) and bremsstrahlung ($\rm R$) interactions. 

The standard inverse function technique was applied in order to
simulate the energy loss $\Delta T$ according to the two-dimensional
distribution $f(\mu,\lambda)$ (Eqn. (1)). For that one needs to
compute the tables of the function $\tilde \lambda=\tilde
\lambda(\mu,\alpha)$, which is inverse to the integral distribution 
\begin{equation}
F(\mu,\lambda)=\int_{-\infty}^{\lambda} 
         f({\mu,\tilde\lambda}) d\tilde\lambda
\end{equation}
These tables can be found, e.g., in Akimov et.al. (1981) where
$\tilde \lambda=\tilde \lambda(\mu,\alpha)$ was tabulated over a
two-dimensional lattice with the steps of 0.025 and 0.01 over $\alpha$
and $\mu$, respectively. 
First, it is necessary to generate the random number 
$\alpha$ uniformly distributed in the interval (0,1) and
calculate parameter $\mu$ according to Eqn. (1).
After that one should interpolate $\lambda=
\lambda(\mu,\alpha)$ using the above mentioned tables and finally 
determine $\Delta T$ as 
\begin{equation}      
\Delta T = \beta_I\Delta l+\lambda Q_I
\end{equation} 
where
\begin{eqnarray}
\lambda= 
\left\{ 
\begin{array}{l}
\mu^{1/2}\cdot \tilde\lambda, \quad \mu < 0.1; \\
\mu\cdot (\tilde\lambda+\ln\mu), \quad \mu \ge 0.1 
\end{array} \right. 
\end{eqnarray}
Formula (5) was derived using two asymptotics of initial distribution 
(1) for $\mu=0$ and $\mu=\infty$, which were sewn up at $\mu = 0.1$.
The subroutine ELSH makes calculations according to these algorithms. 

\subsubsection{Angular deflection}

We simulated the angle of multiple scattering using the Moliere theory
(Moliere 1948, Bethe 1953). In addition we have improved the Moliere distribution 
by taking into account the energy losses at the multiple scattering
segment. This distribution may be described as follows:

\begin{equation}
P(\Theta)\Theta d\Theta=
\left[f_0(\tilde\Theta)+\frac{f_1(\tilde\Theta)}{B}+
\frac{f_2(\tilde\Theta)}{B^2}\right]\tilde\Theta d\tilde\Theta,
\label{Pteta}
\end{equation}
where
\begin{eqnarray}
\tilde\Theta^2=\frac{\Theta^2}{\tilde\chi_c^2}\cdot B, \nonumber \\
\tilde\chi_{c}^2=
\int_{T_o}^{T_1}\chi_c^2(T)/\beta_l (T)dT, \nonumber \\
T_o =  T_1-\beta_{l}(T_1)\Delta l .
\label{XI}
\end{eqnarray}
Parameter $\rm B$ was determined by resolving the transcendental equation
$B-\ln B=\ln(0.856 (\tilde\chi_c/\chi_a)^2)$. Functions $f_n (n=0,1,2)$ 
were calculated as
\begin{equation}
f_n(\tilde\Theta)=\int_{0}^{\infty}(\frac{u^2 ln u}{4})^n e^{-\frac{u^2}{4}}
J_o(\tilde\Theta u) u du.
\end{equation}
Quantities $\chi_c$ and $\chi_a$  are closely related to the 
differential cross-section for Coulomb scattering (see 
formula (10) for definition of the Coulomb cross-section).    

For a fixed segment length $\rm \Delta l$ and kinetic energy of a particle 
$\rm T$ one can determine in a few iterations the parameter $\rm B$, which 
defines the shape of the distribution $\rm P(\Theta)$. Such calculations 
were done by use of subroutine ANGMH. 

\subsubsection{Space displacement}

At the multiple scattering segments we simulated the 
longitudinal displacement $\rm Z$ of a charged particle. 
For that we used the Yang-Spenser 
distribution (Yang 1951, Spencer \& Coune 1962). This distribution 
was adapted for the air shower simulations by Plyasheshnikov \& 
Kolchuzhkin (1975). This distribution may be represented by the 
following expression
\begin{eqnarray}
P(Z)dZ=- e^{\nu}f(\nu,\xi)d\xi \nonumber \\ 
f(\nu,\xi)= \frac{1}{2\pi i}\int_{c-i\infty}^{c+i\infty}u^{1/2} cosch(u^{1/2})
e^q du \nonumber \\
q = \xi u -\nu u^{1/2}ctg(u^{1/2}),
\end{eqnarray}
where
\begin{eqnarray}
\nu=\frac{1-cos\Theta}{\eta\Delta l}, \,\,
\xi=\frac{\Delta l - Z}{\eta\Delta l^2}, \nonumber \\
\eta=2\pi\int_{-1}^{+1}W_c(T_1, \theta)(1-cos\theta)d\cos\theta \nonumber \\
W_c(T,\theta) = 2 \chi_c^2 /(\theta^2+\chi_a^2)^2.
\end{eqnarray}
Here $\rm W_c(T,\theta)$ is a cross-section of the Coulomb scattering which 
determines parameters $\rm \chi_a$ and $\chi_c$. 

To simulate the longitudinal displacement $Z$ we use  
approach similar to that used in Section 2.1.1 for 
simulation of the energy loss $\Delta T$. It is based
on the interpolation of the two-dimensional function $\xi=
\xi(\nu,\alpha)$, which is the inverse function to the integral 
distribution
\begin{equation}
F(\nu,\xi)=\int_{0}^{\xi} f(\nu,\tilde\xi) 
                                      d\tilde\xi
\end{equation}
One can find these tables, e.g., in Akimov et.al (1981).
This approach includes (i) sampling of the random number $\alpha$, 
(ii) calculation of $\nu$ on the basis of quantities $T_1$ 
and $\Delta l$, (iii) determination of $\xi$ on the basis
of the two-dimensional interpolation using the above mentioned tables
and, finally, calculation of $Z$ according to the formula  
\begin{equation}
Z=\Delta l - \xi\eta\Delta l^2.
\end{equation}

For the lateral displacement of a charged particles at the multiple
scattering segment we defined the corresponding probability distribution
using Fermi formula (see e.g., Kolchuzhkin \& Plyasheshnikov 1975).
This distribution was derived by solving the transport equations
in the Focker-Planck approximation where the collision integral 
corresponding to the Coulomb scattering was replaced by a second order
differential operator (see e.g., Kolchuzhkin \& Uchaikin 1978). 
For the fixed angle of multiple-scattering $\vec\Theta=$ 
$(\Theta_x,\Theta_y)=$ $(\Theta\cos\varphi,\Theta\sin\varphi)$ 
this distribution was as follows
\begin{eqnarray}
P(\vec\psi)
=\frac{6}{\pi\gamma}\exp\left[-\frac{6}{\gamma}
(\vec\psi-\vec\Theta/2)^2\right], \nonumber \\  
\vec\psi=\vec\rho/\Delta l,
\label{Ppsi}
\end{eqnarray}
where the parameter $\gamma=\eta\Delta l$ defines the width of the 
distribution.
As was shown by Kolchuzhkin \& Plyasheshnikov (1975) more accurate 
numerical solution of the transport equations reduces by a factor 
of 1.5 the width of the final distribution. This difference may be 
corrected by introducing another definition for the parameter $\gamma$: 
\begin{equation}
\gamma= \chi_c^2 \cdot \tilde B.
\end{equation}
where $\rm \tilde B$ is determined from 
$\tilde B - ln \tilde B = ln((\tilde\chi_c/\chi_a)^2/1.80)$.
The complete procedure for charged particle transport in multi-dimensional 
phase space of energy, angular and space coordinates was defined by 
a few parameters: two threshold energies $\rm Q_I$, $\rm Q_R$, and the 
length of a multiple-scattering segment $\rm \Delta l$. The extensive 
test calculations using the ALTAI code revealed the optimum values of 
parameters which allow to perform rather fast shower simulations without 
introducing systematic errors. Thus we used the following parameters for the 
calculational procedure: $\rm Q_I \simeq 0.5\, T_o$, $\rm Q_R\simeq 0.1\, T_o$, 
where $\rm T_o\simeq 20$~MeV is the threshold energy for \v{C}erenkov 
light emission by cascade electrons in air. 


Apparently the length of a multiple scattering segment, $\rm \Delta l$, 
is one of the basic parameters of this method. By use of rather small 
segments one can reduce systematic error introduced by approximations of 
analytical solutions for the phase transformations at the multiple 
scattering segment. On the other hand this may slow down the procedure of 
shower simulation. We found that the optimum length of the multiple-scattering 
segment is within the range of $\rm \sim 1-5 \, gr/cm^2$. Very accurate 
analytical solutions of a multiple-scattering process for a $e^-,e^+$ 
transport allow us to use in calculations such segment length without 
introducing systematic errors in three-dimensional shower development. 

In comparison to other computational codes we used more accurate distributions 
derived analytically from the multiple scattering theory. On the contrary 
the standard approach used for example in EGS-IV code does not include 
fluctuations of the energy losses at the multiple-scattering segment. 
In addition, in the EGS-IV code the lateral displacement of charged particle at 
the multiple-scattering segment is taken as $\vec\rho=0$ and correspondingly 
the longitudinal displacement is of $z=\Delta l$. All this necessitates a small 
segment length and makes significantly more time consuming the shower 
simulations. Thus for the simulations of \v{C}erenkov light from the 
TeV air shower ALTAI code is faster by as much as a few times compared 
with the EGS-IV. 
  
\subsection{Emission of \v{C}erenkov Light}

A charged particle ($e,\mu$) in an air shower can emit \v{C}erenkov light 
when its energy exceeds a certain threshold energy $\rm E>E_{th}$. 
The threshold energy is determined as (Frank \& Tamm 1937)  
\begin{equation}
E_{th}=\frac{m_o c^2}{\sqrt {2(n-1)}}
\end{equation}
where $\rm n$ is the refraction index in air, $\rm m_o$ is the particle mass.
Thus at sea level the energy threshold is  $E_{th}\simeq$20~MeV for 
electron and $\simeq$4~GeV for muons. To describe the altitude dependence of 
the refraction 
index we used the following expression (Beliaev et al. 1980)
\begin{equation}
n=1+\eta, \quad \eta(H)=\eta_o/\rho_o\cdot\rho(H)
\end{equation}  
where $\eta_o=2.9\cdot 10^{-4}$, $\rho_o=1.22\cdot 10^{-3}\, \rm g/cm^3$, 
and $\rho(H)$ is the air density at a height H above the sea. The model of 
a standard atmosphere (Elterman 1968) was used in the simulations.

\begin{figure}[t]
\includegraphics[width=1.00\linewidth]{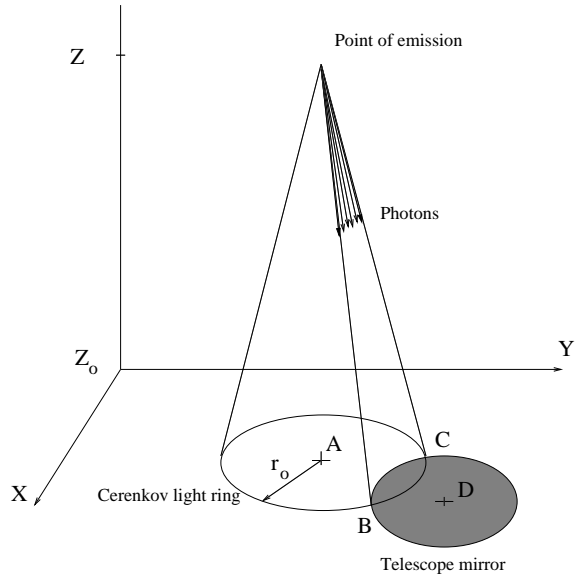}
\caption{Geometry of propagation of \v{C}erenkov light photons.}  
\label{fig:2}
\end{figure}

According to Frank \& Tamm (1937) the mean number of \v{C}erenkon photons 
emitted in a 1~cm pathlength by electron is described as
\begin{eqnarray}
\frac{dQ}{dl} = 2\pi \alpha (\frac{1}{\lambda_1}-\frac{1}{\lambda_2})
\sin^2\theta_c \nonumber \\ 
\sin^2\theta_c=1-\frac{1}{(1-(mc^2/E)^2)n^2}
\end{eqnarray}
where $\alpha$ is a fine structure constant $\alpha=1/137$ and 
$\theta_c\simeq 1^o$ is the opening angle of the \v{C}erenkov light cone.
The spectral region of emission is defined by wavelengths $\lambda_1$ and 
$\lambda_2$. In simulations we included sampling of the \v{C}erenkov light 
attenuation in the atmosphere due to the Raleigh scattering, ozone and aerosol 
absorption. The cross-sections for these processes were calculated using the 
data of Driscoll \& Vaughan (1978). 

In the shower simulation procedure (PARCHR, GSTCHR) (Konopelko 1990), 
we first define the total number of 
\v{C}erenkov photons $\rm <n_{ph}>$ emitted at the multiple scattering segment 
$\rm \Delta l$. As was discussed above we used rather small multiple scattering 
segments. Therefore, we can assume that all \v{C}erenkov photons are 
emitted from the center of the multiple scattering segment (see Figure~1).
To a rather good approximation an intersection of the \v{C}erenkov light cone 
with the observation plane forms a circle of a radius $\rm r_c$ which can
be calculated as 
\begin{equation}
r_c=(Z-Z_{obs}) \cdot \theta_c 
\end{equation}
where $Z$ and $Z_{obs}$ are the height of \v{C}erenkov light emission and the 
height of the observation level above the sea, respectively. The number of 
photons hitting the telescope mirror can be calculated as 
\begin{equation}
<\tilde n_{ph}> =\frac{\Delta L}{2\pi r_c}(1-p(Z,Z_{obz}))<n_{ph}>
\end{equation} 
where $\Delta L$ is length of the circle arc between points B and C (see Figure~2).
$\rm p(Z,Z_{obs})$ is a probability of the \v{C}erenkov light attenuation 
in the atmosphere along the way of a photon propagation. For specific 
response functions of the telescope camera one may calculate corresponding average 
number of photoelectrons as $\rm <n_{ph.e.}> = \xi <\tilde n_{ph}>$, where $\xi$ is 
a photon-to-photoelectrons conversion efficiency. Finally one can use the Poisson 
distribution in order to simulate the random number of photons (photoelectrons) 
hitting the telescope. The photons reaching the telescope mirror were uniformly 
distributed over the circle arc BC (see Figure~2).

In the latest version of the ALTAI code we save all, or a certain fraction, of  
\v{C}erenkov photons hitting the telescope. Each photon has a set of 6 variables, 
$\vec \upsilon (x, y, \theta_x, \theta_y, z_o, t)$. $\rm z_o$ is a height of the 
photon emission (in $\rm g/cm^2$) and $t$ is a time of photon arrival to the 
detector. Such database of the photon is used for detailed simulations of 
the telescope camera response (Hemberger 1998).

\begin{table*}[t]
\setlength{\tabcolsep}{1.5pc}
\newlength{\digitwidth} \settowidth{\digitwidth}{\rm 0}
\catcode`?=\active \def?{\kern\digitwidth}
\caption{Mean number of fragments created by different nuclei
(Murzin 1988).}
\label{tab:1}
\begin{tabular*}{\textwidth}{@{}l@{\extracolsep{\fill}}rrrr}
\hline
Primaries $\mid$ Secondaries & $Z>10$   & C,N,O,F  & Li, Be, B &  He   \\ \hline
$Z>10$                  &  0.17    &     0.29 &    0.26   &  1.34 \\ 
C,N,O,F                 &   --     &     0.11 &    0.24   &  1.00 \\ 
Li,Be,B                 &   --     &     --   &    0.15   &  0.51 \\  \hline
\end{tabular*}
\end{table*}

\section{Hadron-nuclei cascade}

For simulations of the hadron-nuclei cascade we used the phenomenological model 
of hadron interactions (Konopelko 1990) which, in the most part, is
based on the available accelerator 
data. The energy spectra of secondary hadrons generated in a 
$pA$ interactions 
were approximated by use of the radial scaling model (Hillas 1979). 
In this approach the energy spectra may be presented as 
\begin{equation}
xdN/dx=F_{(p\rightarrow q)}(x) H_q(x,E), x=E/E_o
\label{mod1}
\end{equation}
where $E_o,E$ are the energies of the primary and secondary particles, 
respectively. Indeces $p,q$ denote the type of primary and secondary particle
($p,n,\pi^+,\pi^-,\pi^o$). The basic functions are given below

\begin{eqnarray}
F_{(p,n)\rightarrow \pi^{\pm}}= \nonumber \\
F_{(p,n)\rightarrow \pi^{0}}= \nonumber \\
1.22(1-x)^{3.5}+0.98e^{-18x}, \nonumber \\ 
F_{\pi^\pm\rightarrow\pi^{0}}=1.3(1+\frac{x}{0.45})^{-3},\nonumber \\
F_{\pi^\pm\rightarrow\pi^{\pm}}=F_{\pi^\pm\rightarrow\pi^{0}}+0.57e^{4(x-1)}, \nonumber \\
H_{(p,n)\rightarrow(\pi^\pm,\pi^o)}=[1+\frac{0.4}{E+0.14}]^{-1}, \nonumber \\
H_{\pi^\pm\rightarrow(\pi^\pm\pi^o)}=1-0.88e^{-1.8x}
\end{eqnarray}

For the total cross-sections of inelastic hadron interactions we used the data 
given by Shabelski (1986,1987) in the following form 
\begin{eqnarray}
\sigma_{\pi^\pm}=1.2\sigma_{(p,n)}, \nonumber \\
\sigma_{(p,n)}=\left\{ 
\begin{array}{l}
258 \,\,E<E_1, \\
273(1+3.29\cdot 10^{-2}y+ \\
3.88\cdot 10^{-3}y^2)\,\,E_1 < E < E_2, \\
258(1+0.07y)\,\,  E > E_2 \\ 
\end{array} \right. \nonumber \\
E_1 = 100 \, GeV, E_2 = 10^4\, GeV, \nonumber \\ 
y=ln(E/10^3\,GeV) 
\label{mod3}
\end{eqnarray}
The cross-sections from Eqn. (22) are measured in mBarn and correspond 
to the particle interaction with air. We assume that the mean value of
atomic number for air nuclei is of 14.4.

We used a simplified geometrical representation (Murzin 1988) for 
the total cross-section of the nucleus-nucleus interactions as follows
\begin{eqnarray}
\sigma_{A} = R^2 \sigma_N, \nonumber \\
R=\frac{A^{1/3}+A_0^{1/3} - b}{1 + A_0^{1/3}-b}, 
\end{eqnarray}
where $A_0$ is the effective atomic mass of air ($A_o = 14.4$) and $b$ 
is the effective radius of nuclei overlapping zone ($b=1.17$). 
$\sigma_N$ defines the cross-section of hadron interaction with air.

The transverse momenta of secondary hadrons were calculated according 
to the distribution given by Ranft (1972) 
\begin{equation}
f_{(p,n,\pi^\pm)}(p_{\perp})=p_{\perp}\frac{e^{-Bp_{\perp}^2}+Ce^{-Dp_{\perp}}}
{\frac{1}{2D}+\frac{C}{D^2}}
\label{mod4}
\end{equation}
where the corresponding parameters (B,C,D) were defined for the interactions
with air as described by Ranft et al.~(1972).
$p_{\perp}$ is measured in GeV/c. 

The ALTAI code was developed for the simulations in the energy range relevant 
for the very high energy $\gamma$-ray astronomy $\rm E \leq 50\, TeV$. In this 
energy range the mean number of kaons ($\rm K^\pm,K^o$) produced in hadron-nuclei 
interactions is very small compared with the number of emitted pions 
($\rm \pi^\pm,\pi^o$). Besides this, in this energy region kaons and pions exhibit 
similar properties of inelastic interactions (see, e.g., Grishin 1982). Thus in 
the shower simulations for this restricted energy range we can exclude the kaon 
production process by introducing appropriate corrections for the probabilities of 
pion production in the $(p,n,\pi^\pm)\rightarrow X$ inelastic collisions. 

For each inelastic hadron interaction (MULTIP) we sampled first the energy of the 
so-called {\it leading} particle ($p,n,\pi^\pm$). This particle carries away the 
bulk of the primary energy. We assume that the energy of a leading particle E  
is uniformly distributed within the range $(0,E_0)$ where $E_0$ is the energy of 
a primary interacting particle. Note that the type of the leading particle was 
selected randomly amongst proton and neutron for interaction $(p,n)+A \rightarrow (p,n)$, 
and amongst charged pion or neutral pion for interaction $(\pi^\pm,\pi^o)+A \rightarrow 
(\pi^\pm,\pi^o)$. Thus the total inelasticity 
coefficient was determined as $\kappa=1-E/E_0$. At the next step we sampled 
multiple production of other secondary particles ($\pi^\pm,\pi^o$). The type of 
secondary particle was randomized assuming that $\pi^o$ pions forms, on average, 
one third of all pions generated in hadron interactions. The energy of secondary 
particle was simulated using the spectra given by Eqn. (\ref{mod1}). We have 
completed the production of secondary particles when their total energy exceeds the energy 
of a primary particle. The energy of the last simulated particle was renormalized 
in order to conserve the total energy in each inelastic interaction. 

The transverse momentum of the secondaries was simulated using Eqn. (\ref{mod4}). 
By the renormalization of the transverse momentum of leading particles we 
allowed to fulfill the momentum conservation for each hadron interaction without 
distortion of stochastic properties of a multi-particle production process. 
The test calculations have shown that such an approach describes very well the 
initial inclusive spectra of the secondary particles if the number of secondaries is 
relatively high ($\gg 1$). Note that as almost identical algorithm was developed and 
tested by Barashenkov \& Toneev (1972).

Note that we have included in the code the propagation of a single muon generated 
in the hadron-nuclei cascade due to the $\pi^\pm$ decay (SGG,TRMUON). The emission 
of the \v{C}erenkov light from the muons was included in simulations according to the 
scheme described above.   

\section{Nucleus-nucleus interactions}

We have implemented in the ALTAI code a model of independent nucleon interactions 
of colliding nuclei with nuclei fragmentation included (TAFRAC, TRAN, FRAG, 
OVERLAP). In this approach we assume that all nucleons of the projectile 
nucleus have the same energy determined as $E_0/A_p$. $E_0$ and $A_p$ are the 
energy and atomic number of projectile nucleus, respectively. All nucleons of the 
projectile nucleus which overlap with the target nucleus interacted independently 
with each other. The non-overlapping part of the projectile nucleus decayed into 
individual nucleons and heavier fragments. The energy of fragment with the atomic 
number $A$ is defined as $E_f=(E_0 \cdot A)/A_p$. We simulated a random number of 
a fragments according to the probabilities of different channels summarized in 
Table~1. 

\section{Comparison with other codes and data}

An overview of Monte Carlo results on lateral, temporal, and angular characteristics 
of the \v{C}erenkov light in air showers of 10~GeV - 1~TeV was recently given by 
Konopelko (1997). Most of the \v{C}erenkov light characteristics calculated with 
the ALTAI code reproduce rather well the results obtained with the MOCCA code (Hillas 1996). 
The {\it stereoscopic} observations of BL Lac object Mkn~501 in 1997 with the 
HEGRA system of imaging air \v{C}erenkov telescopes allowed the 
first measurements  
of the parameters of the \v{C}erenkov light images from the $\gamma$-ray-induced 
air showers. The HEGRA data are in excellent agreement with the results of Monte Carlo 
simulations obtained with ALTAI code (Aharonian et al. 1999a).
Using the Mkn~501 data sample comprising 38,000 $\gamma$-ray events we have tested 
in great detail the parameters of \v{C}erenkov light image orientation and shape 
(Konopelko et al. 1999a) (see Figure~3) as predicted by the simulations. 

\begin{figure}[t]
\includegraphics[width=1.00\linewidth]{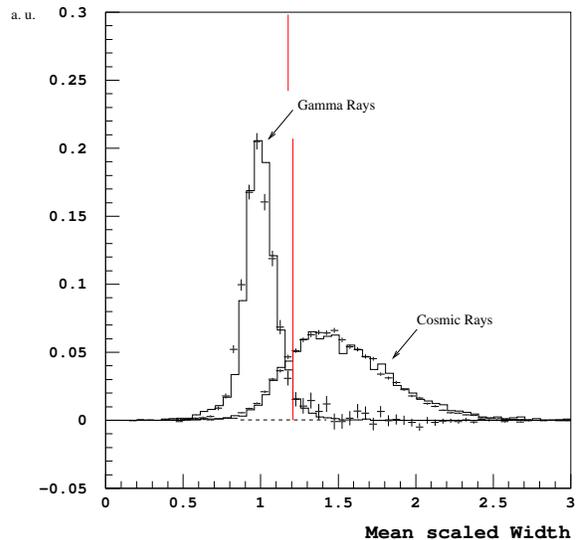}
\caption{Distribution of a shape parameter (mean scaled Width) of \v{C}erenkov light 
images simulated with ALTAI code (histograms) and recorded with HEGRA system of 
IACTs (crosses).}  
\label{fig:1}
\end{figure}

\begin{figure}[t]
\includegraphics[width=0.98\linewidth]{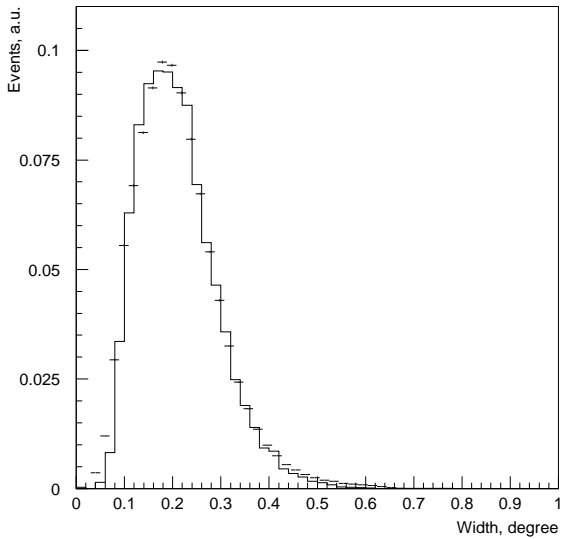}
\caption{Distribution of image shape parameter Width for the ALTAI simulated events (histogram)
and cosmic ray images recorded with a 10~m Whipple IACT (crosses) (Konopelko 1999b).}  
\label{fig:1}
\end{figure}
 
The HEGRA stereoscopic system of 5 IACTs was used for the measurements of the lateral 
distribution of the \v{C}erenkov light in the $\gamma$-ray-induced air showers. These 
measurements have been compared again with the Monte Carlo simulations using the ALTAI code 
(Aharonian~et~al. (1998)). The simulations using the ALTAI code reproduce very well the 
measured lateral distributions of the \v{C}erenkov light. Note that the measured shape of 
the \v{C}erenkov light lateral distributions is almost independent of the detector 
simulation procedure but strongly depends on the development in space of a multi-TeV 
$\gamma$-ray shower in the atmosphere.

We made several comparisons of the shape and size (total number of photoelectrons) 
of the \v{C}erenkov light images for the proton- and nuclei-induced air shower 
calculated with the ALTAI and CORSIKA codes (Heck et al. 1997). The CORSIKA code 
was used with the HDPM model of the proton-nuclei interactions. 
This model is based on supercollider data and describes rather well the proton-nuclei 
air showers of energy below 10~TeV. In addition, the HDPM model of CORSIKA allows  
to perform relatively fast calculations with respect to other more modern shower generators, 
e.g. VENUS/QGSJET. Despite the different algorithms and schemes in the two codes the results 
appeared to be almost identical (Plyasheshnikov~et~al. 1997). The simulations with the ALTAI and 
CORSIKA codes well reproduce each other. Note that simulations with the ALTAI code are essentially 
less time consuming. The HEGRA data for the cosmic ray air showers were compared with the 
ALTAI simulations. Good agreement between simulations and data provided a precise measurement 
of the cosmic ray proton spectrum in the energy range $\sim$1-5~TeV (Aharonian et al. 1999b).

We show in Figure~3 the distribution of so-called {\it mean scaled Width parameter} 
for primary $\gamma$-rays and cosmic rays extracted from the HEGRA data as well as from 
the Monte Carlo simulations. The simulations fit very well the data. Recently, we 
compared the relevant results of the ALTAI simulations with the  
cosmic ray data taken with a 10~m Whipple imaging  air \v{C}erenkov telescope during 
the 1995/1996 Crab~Nebula observations (Konopelko 1999b). For these simulations we have 
used the standard detector response functions offered by the Whipple collaboration. 
The simulations reproduced very well the shape of recorded \v{C}erenkov light images 
(see Figure~4). 

As mentioned above the upper energy of the shower simulations is about 50~TeV. It is 
limited by the simplified phenomenological model of the hadronic cascade. Although 
in the case of pure electromagnetic showers one can extend simulations using the ALTAI 
code to much higher energies without breaking any model restrictions
such extension may need the introduction of a number of changes into the code for the 
correct treatment of a larger number of particles.

\section{Summary}

Here we have presented a detailed description of the numeric code ALTAI, which was 
developed for simulations of \v{C}erenkov light from  
extensive air showers. This code allows to make very fast and accurate calculations of 
the response of the ground-based \v{C}erenkov detectors used in VHE $\gamma$-ray 
astronomy. 
Although the code was designed for calculating the parameters and 
characteristics of the imaging air \v{C}erenkov telescopes it can be rearranged with 
minor changes in order to simulate the responses of showerfront sampling experiments 
like MILAGRO, or Tibet AS-$\gamma$. 

We tested our computational code against the data taken with two currently 
operating detectors, the HEGRA system of imaging air \v{C}erenkov telescopes and 
the state-of-the-art single 10~m Whipple telescope. These comparisons have proven the high 
precision of the simulations. The computational code ALTAI is an effective tool for 
producing the simulated data for VHE $\gamma$-ray astronomy.   
Note that the forthcoming imaging \v{C}erenkov detectors (CANGAROO~IV, H.E.S.S., 
MAGIC, VERITAS) will need a large amount of simulated data. Therefore the performance 
of these instruments may benefit from the use of the ALTAI code. 

\section{Acknowledgments}

The work for the ALTAI code was primarily initiated and substantially advanced at the 
Tomsk Technological University, Tomsk, Russia, where the major algorithms of the shower 
simulation scheme were developed. Authors thank Prof. A.M. Kolchuzhkin for a significant 
contribution for all these studies. In the most part the ALTAI code was developed and 
programmed at the Altai State University, Barnaul, Russia. The authors thank 
Dr K.V. Vorobjev, Prof A.M. Lagutin, Dr V.A. Litvinov and Prof V.V. Uchaikin 
for their valuable input and support of this activity. The computational code ALTAI was used and 
further developed at the Max-Planck-Institut f\"ur Kernphysik, Heidelberg. The authors would 
like to acknowledge the contribution and support of all members of the Heidelberg group 
in particular Prof F. Aharonian, Dr M. Hemberger, Prof W. Hofmann, J. Kettler and 
Prof H.J. V\"olk. AKK thanks Prof T.C. Weekes for support and supervision of a 
short-term project at the University of Arizona, Tucson, and at the Whipple Observatory, 
Harvard-Smithsonian Center for Astrophysics, Amado.    

We are grateful to an anonymous referee for detailed and helpful comments.

\section{Availability}

The ALTAI code is written in FORTRAN and may be easily installed at any computer platforms 
maintaining FORTRAN compiler. Regarding the availability of the code contact:
alexander.konopelko@mpg.mpi-hd.de

\end{document}